\documentclass[10pt,conference]{IEEEtran}
\IEEEoverridecommandlockouts
\usepackage{amsmath,amssymb,amsfonts}
\usepackage{algorithmic}
\usepackage{graphicx}
\usepackage{textcomp}
\usepackage{xcolor}
\usepackage{subcaption}
\usepackage{subfig}
\usepackage{tikz}
\usepackage{pgfplots}
\usepackage{cleveref}
\usepackage[numbers,sort&compress]{natbib}

\usetikzlibrary{positioning,arrows.meta}

\def\BibTeX{{\rm B\kern-.05em{\sc i\kern-.025em b}\kern-.08em
    T\kern-.1667em\lower.7ex\hbox{E}\kern-.125emX}}
\begin{document}

\newcommand{\SJ}[1]{ {\color{blue}{SJ:#1}} }
\newcommand{\hk}[1]{ {\color{red}{HK:#1}} }
\newcommand{\rishabh}[1]{ {\color{green}{rishabh:#1}} }
\newcommand{\Mark}[1]{ {\color{teal}{Mark:#1}} } 

\newcommand*\circled[1]{\tikz[baseline=(char.base)]{
            \node[shape=circle,draw,inner sep=1pt] (char) {#1};}}

\title{Hardware vs. Software Implementation of  Warp-Level Features in Vortex RISC-V GPU
}


\author{
    \IEEEauthorblockN{
        Huanzhi Pu\IEEEauthorrefmark{1}, 
        Rishabh Ravi\IEEEauthorrefmark{2}, 
        Shinnung Jeong\IEEEauthorrefmark{1}, 
        Udit Subramanya\IEEEauthorrefmark{1}
    } 
    \IEEEauthorblockN{
        Euijun Chung\IEEEauthorrefmark{1}, 
        Jisheng Zhao\IEEEauthorrefmark{1}, 
        Chihyo Ahn\IEEEauthorrefmark{1},
        Hyesoon Kim\IEEEauthorrefmark{1}
    }   
    \\    
    \IEEEauthorblockA{\IEEEauthorrefmark{1}College of Computing, Georgia Institute of Technology, Atlanta, USA \\ 
    \{hpu8, shinnung, usubramanya3, echung67, jisheng.zhao, ahnch, hyesoon\}@gatech.edu}
  \IEEEauthorblockA{\IEEEauthorrefmark{2}Department of Electrical Engineering, Indian Institute of Technology Bombay, Mumbai, India \\ 
    200260041@iitb.ac.in}
    
}
\maketitle

\crefformat{section}{\S#2#1#3}
\crefname{section}{Section}{Sections}
\crefname{figure}{Figure}{Figures}
\crefname{table}{Table}{Tables}
\crefname{alg}{Algorithm}{Algorithms}
\crefname{equation}{Equation}{Equations}
\creflabelformat{equation}{#2\textup{#1}#3}
\crefname{appendix}{Appendix}{Appendices}
\crefname{algorithm}{Algorithm}{Algorithms}
\crefname{lstnumber}{line}{lines}
\crefname{lstnumber}{Line}{Lines}
\crefname{lstlisting}{listing}{listings}
\Crefname{lstlisting}{Listing}{Listings}

\begin{abstract}
RISC-V GPUs present a promising path for supporting GPU applications. Traditionally, GPUs achieve high efficiency through the SPMD (Single Program Multiple Data) programming model. However, modern GPU programming increasingly relies on warp-level features, which diverge from the conventional SPMD paradigm. In this paper, we explore how RISC-V GPUs can support these warp-level features both through hardware implementation and via software-only approaches. 
Our evaluation shows that a hardware implementation achieves up to 4 times geomean IPC speedup in microbenchmarks, while software-based solutions provide a viable alternative for area-constrained scenarios. 

\end{abstract}

\begin{IEEEkeywords}
GPU, Warp-level features, Microarchitecture, Code Optimization
\end{IEEEkeywords}

\section{Introduction} 
\label{sec:intro}

In recent years, GPU programming models have expanded the scope of GPU programming by enabling fine-grained parallelism through warp-level features. This expansion allows GPU to operate within the SPMD programming model while diverging from the conventional SPMD paradigm by fine-grained thread control.
In particular, CUDA~\cite{nvidia_cuda,cuda_toolkit}, one of the most widely used GPU programming models, introduces warp-level features such as cooperative groups and warp-level functions to facilitate fine-grained thread control and synchronization. These warp-level features provide abstractions that go beyond fixed-size granularity and synchronization, allowing for more complex code and reducing the need for block-level synchronization barriers

Therefore, supporting warp-level features in RISC-V GPUs~\cite{vortex, vortex_first_2020, vortex:skybox, andryc2013flexgrip, kadi2016, duarte2017scratch, ma2021} can provide opportunities by enhancing their generality and applicability. While RISC-V GPUs, such as the Vortex RISC-V GPU~\cite{vortex, vortex_first_2020, vortex:skybox}, present a promising path for supporting GPU applications with publicly available software and hardware stacks and offer high reconfigurability for diverse GPU hardware features, they lack support for recent high-level features such as warp-level functionality. As a result, they miss the opportunity to explore higher performance by leveraging their reconfigurable hardware features, such as the number of threads and warps, in combination with warp-level features.

To efficiently support warp-level features, we explore implementation methods on Vortex GPUs by considering both software-only and hardware-supported approaches with their own trade-offs. While hardware extensions can offer performance benefits, they also incur additional area costs. Conversely, a software-only approach avoids hardware overhead but requires additional instruction overhead and compiler support. To enable both approaches, we implement these features in Vortex RTL and extend compiler support.

Through a comparative analysis of the two implementation approaches, this paper evaluates how these approaches support warp-level features and adapt to Vortex GPU architectures. The results show that warp-level feature support requires minimal hardware overhead with only a 2\%  increase in logic area, while the performance difference between software and hardware implementations can be as much as 4× in microbenchmarks.


\section{Background}
\label{sec:back}

\subsection{Vortex RISC-V GPU}
\label{sec:back:vortex}

\begin{figure}[t]
  \centering
  \includegraphics[width=0.85\linewidth]{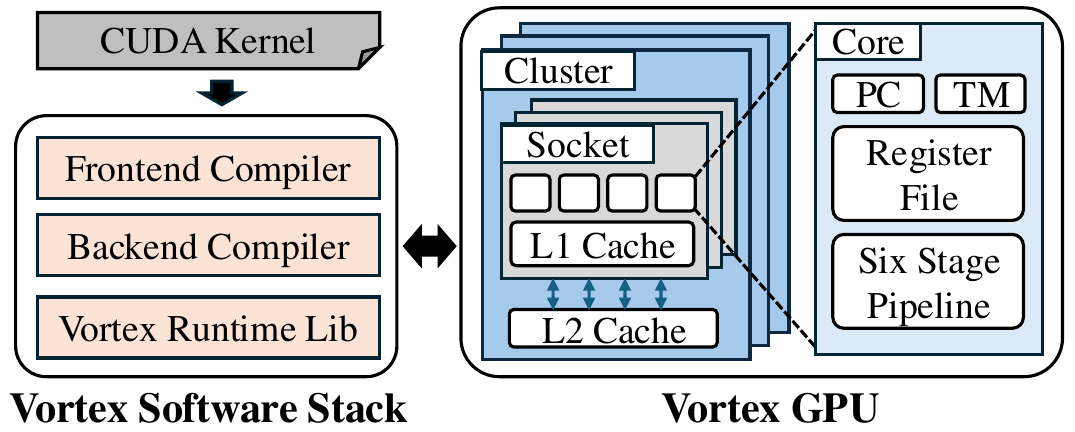}
  \caption{Vortex GPU and software stack}
\label{fig:back:vortex}
\vspace{-3mm}
\end{figure}

Recently, some works~\cite{vortex, vortex_first_2020, vortex:skybox} have proposed a RISC-V GPU called Vortex GPU, which features a highly reconfigurable architecture and a modular and comprehensive software stack. Vortex GPU supports various hardware configurations, such as adjusting the number of warps or cores, while its software stack provides compiler support for diverse programming models such as CUDA~\cite{nvidia_cuda,cuda_toolkit} and OpenCL~\cite{OpenCL}. Since both its hardware and software stack are fully open-source, some works~\cite{vortex:ipdpsw, vortex:sparseweaver,vortex:lmi} extend Vortex GPU to enhance GPU features and explore new optimizations.

\Cref{fig:back:vortex} shows an overview of the Vortex GPU hardware architecture and its software stack. The GPU architecture consists of three hierarchies called Cluster, Socket, and Core. The Cluster and Socket are connected through L1 and L2 caches, and the Core features a six-stage pipeline with its own register file, program counter (PC), and warp scheduler. The software stack consists of two different compilers: a frontend compiler, such as CuPBoP~\cite{han2024cupbop} for CUDA support, and a backend compiler to translate kernels into device kernels consisting of the Vortex Instruction Set Architecture (ISA). Vortex also supports the Vortex runtime library to manage communication between the host CPU and GPU.


\vspace{-1mm}

\subsection{CUDA Warp-Level Features}
\label{sec:back:cuda}



CUDA~\cite{nvidia_cuda,cuda_toolkit} is one of the widely used programming models proposed by Nvidia.  CUDA programming model is structured in a three-level hierarchy: a kernel represents the function executed on the GPU, which is launched as a grid consisting of multiple thread blocks, each containing a group of threads. Threads within a block can collaborate through shared memory and synchronization, while execution follows the SPMD model.
However, the conventional CUDA programming model, designed for the SPMD paradigm, lacks fine-grained thread control and efficient synchronization beyond the block or warp level. These limitations degrade performance in some workloads requiring frequent inter-thread communication or fine-grained parallelism. To address this, CUDA introduces warp-level features, such as warp-level functions and cooperative groups.

\textbf{Warp-Level Functions} enable fast data sharing among threads by exchanging register values and providing fine-grained synchronization within a warp~\cite{cuda_toolkit}. CUDA provides several warp-level primitives, including shuffle, reduce, vote, and match. In particular, the shuffle primitive facilitates value exchange among threads, while the vote primitive aggregates the satisfaction status of a specific condition within a warp. 


\textbf{Cooperative Groups} enable flexible and user-defined hierarchical synchronization and grouping of threads, extending beyond the traditional programming model hierarchy~\cite{cuda_toolkit}.
Originally, CUDA only provided barriers for synchronizing threads within a block, and warp-level synchronization was later introduced. Cooperative Groups provide a safe method to synchronize threads at various granularities within a block and extend synchronization capabilities beyond traditional barriers to improve program efficiency. In addition, Cooperative Groups provide an abstraction for defining and managing thread groups at different levels. The organized interface allows developers to explicitly express thread grouping logic and write safe, maintainable, and efficient code.

Since warp-level features can improve the GPU programming paradigm, supporting these features in Vortex GPU can expand its generality and applicability.
In this paper, we focus on supporting Cooperative Groups and warp-level functions (shuffle and vote) using both software-only and hardware-supported approaches to analyze their efficiency from performance and area perspectives.


\section{Hardware Solution}
\label{sec:hw}




\begin{figure}[t]
  \centering
  \includegraphics[width=0.43\textwidth]{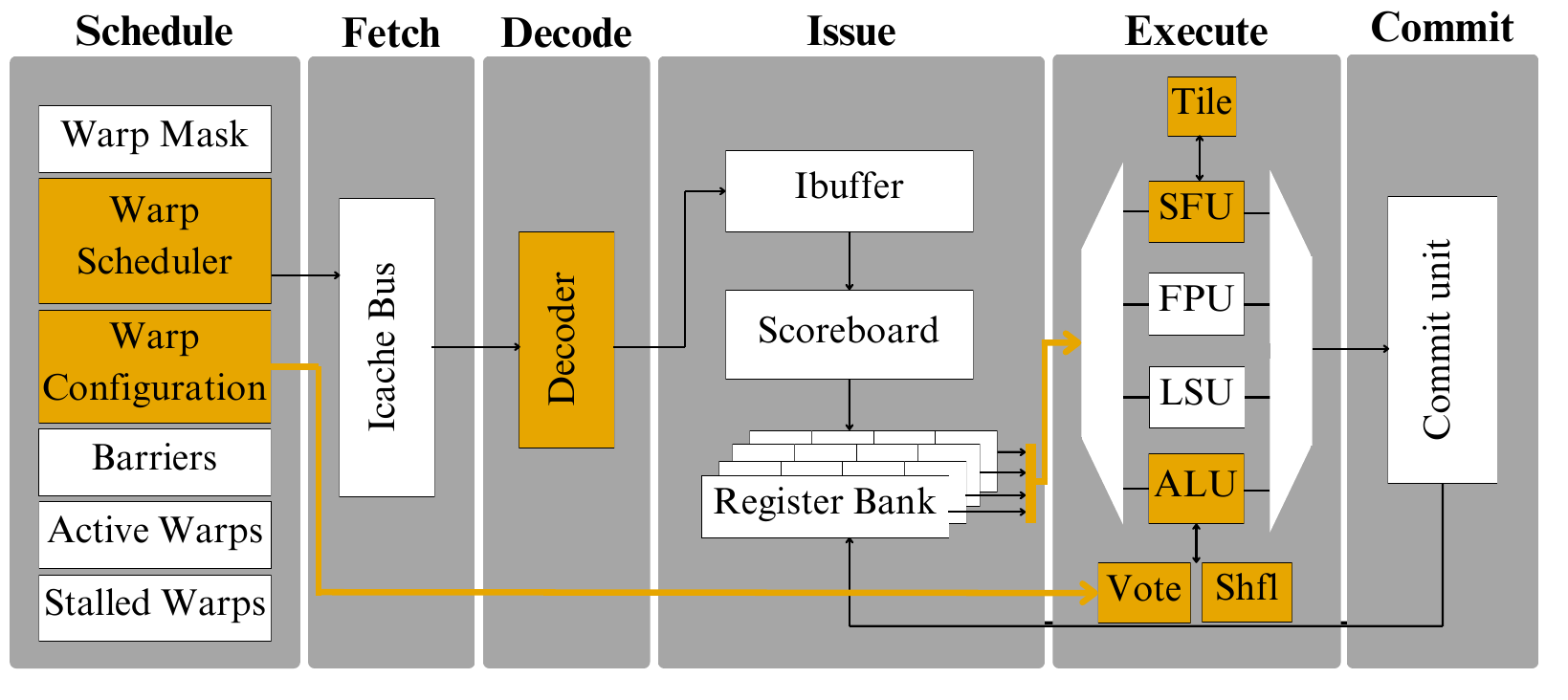}
  \caption{Vortex core architecture. Modified sections (highlighted in yellow) support cooperative groups and warp-level functions}
\label{fig:hw:vortex_extension}
\vspace{-1mm}
\end{figure}

\begin{table}[t]
    \centering
    \footnotesize
    \begin{tabular}{|c|c|c|c|}
        \hline
        \textbf{Operation} & \textbf{IType} & \textbf{Opcode} & \textbf{func}\\
        \hline
        vx\_vote & I & CUSTOM0 & All,Any,Uni,Ballot \\
        \hline
        vx\_shfl & I & CUSTOM1 & Up,Down,Bfly, Idx \\
        \hline
        vx\_tile & R & CUSTOM2 & - \\        \hline
    \end{tabular}
    \caption{Extended RISC-V Instructions for warp level collectives and cooperative groups}
    \vspace{-5mm}
    \label{tab:instructions}
\end{table}

\begin{table}[t]
    \centering
    \footnotesize
    \begin{tabular}{|c|c|c|} \hline
        \textbf{Configuration} & \textbf{Group Mask} & \textbf{size}  \\ \hline
        No groups (default) & 10000000 & 32 \\ \hline
        2 groups- 16 threads & 10001000 & 16 \\ \hline
        4 groups - 8 threads & 10101010 & 8 \\ \hline
        8 groups - 4 threads & 11111111 & 4 \\ \hline
    \end{tabular}
        \caption{Configurations set by {\it tile}}
    \label{tab:config}
    \vspace{-5mm}
\end{table}

\begin{figure}[t]
    \begin{subfigure}[b]{0.23\textwidth}
        \includegraphics[width=\textwidth]{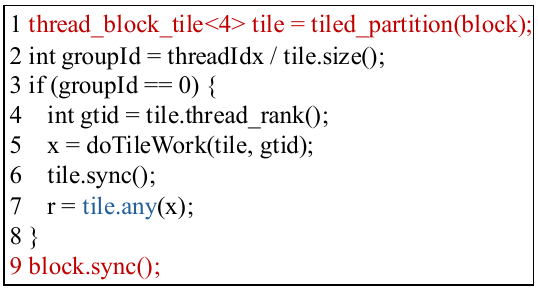}
        \caption{Original kernel}
        \label{fig:original_code}
    \end{subfigure}
    \hfill
    \begin{subfigure}[b]{0.23\textwidth}
        \includegraphics[width=\textwidth]{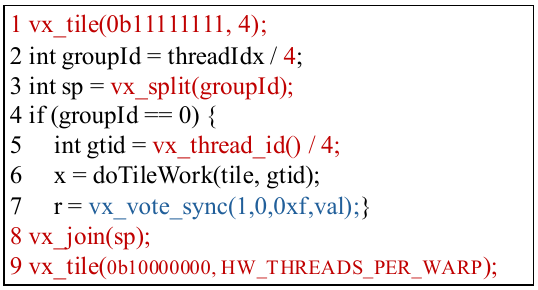}
        \caption{Hardware intrinsics}
        \label{fig:hw_code}
    \end{subfigure}
    \caption{CUDA kernel example and converted Pseudo-code kernel example using hardware intrinsics}
    \vspace{-5mm}
    \label{fig:hw_code_example}
\end{figure}

To support warp-level functions and cooperative groups at the hardware level, we need architectural modifications and ISA extensions. \Cref{fig:hw:vortex_extension} shows the modified Vortex core architecture and \Cref{tab:instructions} presents the introduced ISA.





\textbf{Warp-Level Functions}: 
To support vote and shuffle functions, we modified the ALU to exchange and count register values, respectively, and the decoder to support the new instructions \textit{vx\_vote} and \textit{vx\_shfl}.
As shown in Table I, both the vote and shuffle operations support four different modes, determined by the function field of the instructions. In addition, the immediate field of vote contains the register address that stores the member mask, which is fetched before execution. Similarly, shfl's immediate field includes the lane offset and the register address that stores the clamp value.

\textbf{Cooperative Groups}: 
We implement cooperative groups by modifying the architecture to support a variable warp structure, enabling synchronization across different granularities. Our design is based on the idea that threads within a warp can be synchronized, and synchronization across thread blocks is translated into warp-level synchronization.  By default,warp starts with a fixed size and dynamically reshapes through merges or splits to match the group size when there are cooperative group instructions. 

To support this idea, we introduce the \textit{vx\_tile} instruction, which takes group mask and thread count as operands to differentiate thread execution. For example, \Cref{tab:config} provides configurations for four different thread execution modes when the hardware thread size in the core is 32. When using a cooperative group, the core can be initialized with warps of four threads and dynamically merge them into larger warps with a thread count specified by the user, such as 16 threads.

While merging, Vortex must prevent erroneous execution by assigning the correct register bank to each warp and ensuring accessibility only by that warp. Therefore, we add a cross-bar instead of a multiplexer to ensure data availability at the execution stage. The scheduler with warp configuration manages the execution unit to ensure correct execution by controlling cross-bar. In addition, we modify the execute unit to guarantee correct operations when the output depends on group size. Synchronization is handled either by existing single-warp support or by modifying group sync exit conditions for merged warps, with all changes localized to the scheduling unit.


\textbf{Code Example}: 
\Cref{fig:hw_code_example} shows a code example using our intrinsic functions in the Vortex runtime library. \Cref{fig:original_code} presents a simple CUDA kernel that uses a cooperative group with a tile size of four and warp-level functions with vote operations, and \Cref{fig:hw_code} provides a pseudo-code example using intrinsics. $\textit{tiled\_partition}$ is replaced by $\textit{vx\_tile}$, and $\textit{vote\_any}$ is replaced with $\textit{vx\_vote\_sync}$ with mode 1. The divergent branch is handled using $\textit{vx\_split}$ and $\textit{vx\_join}$.

\section{Software Solution}
\label{sec:sw}


\begin{figure}[t]
    \begin{subfigure}[b]{0.23\textwidth}
        \includegraphics[width=\textwidth]{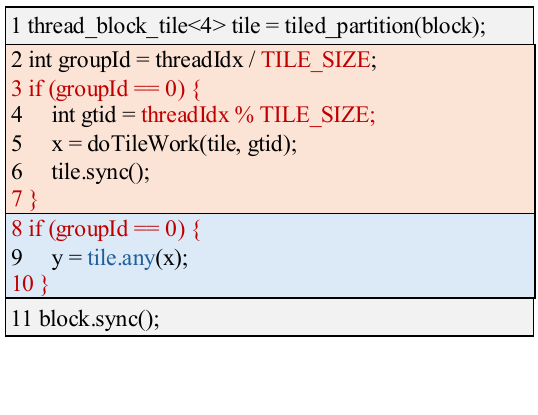}
        \caption{Identified parallel regions}
        \label{fig:sw:pr}
    \end{subfigure}
    \hfill
    \begin{subfigure}[b]{0.23\textwidth}
        \includegraphics[width=\textwidth]{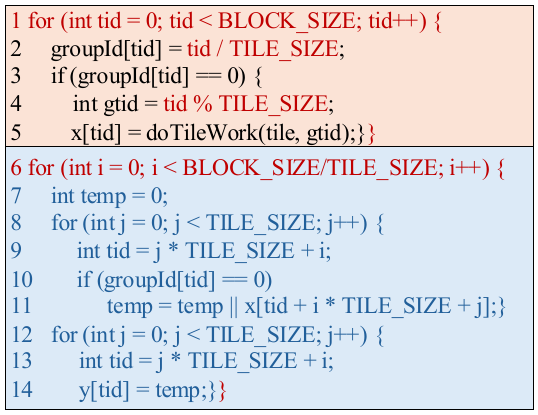}
        \caption{Kernel after PR transformation}
        \label{fig:sw:loop}
    \end{subfigure}
    \caption{Transformed kernel after applying PR transformation in \Cref{fig:original_code} ({\bf SW} solution)}
    \vspace{-1mm}
    \label{fig:sw_code}
\end{figure}

This section proposes an extended parallel region transformation algorithm to support warp-level features without Vortex hardware support. The key idea of parallel region transformation is identifying regions in the GPU kernel that can be executed in parallel (called parallel regions) and applying loop serialization to map software thread blocks onto hardware threads~\cite{han2024cupbop, cox}.
Based on this idea, we extend the parallel region transformation algorithm by incorporating nested loops and divergence control to support warp-level features.

The proposed parallel region transformation algorithm (PR transformation) consists of the following processes:
\circled{1} The frontend compiler identifies parallel regions. The boundaries of the parallel regions are defined by cross-thread operations, including synchronization, block partitioning, warp-level operations, and cooperative group operations.
\circled{2} The compiler performs control structure fission to apply divergence control when if or if-else structures span multiple parallel regions.
For example, after if-else fission is applied, the kernel in \Cref{fig:original_code} is divided into four colored parallel regions, as shown in \Cref{fig:sw:pr}.
\circled{3} After that, the compiler removes parallel regions that contain only synchronization and partitioning functions.  Thus, the gray PRs in \Cref{fig:sw:pr} are removed.
\circled{4} The compiler applies loop serialization, considering a transformation rule for warp-level features in \Cref{sec:sw:rule}.
Generally, each parallel region is converted using a single loop, such as the converted orange block in  \Cref{fig:sw:loop}. However, the compiler uses nested loop serialization for warp-level functions. For example, the compiler applies nested loop serialization for vote functions, as shown in the blue region of \Cref{fig:sw:loop}.
\circled{5} The compiler applies optimizations to handle special variables, including thread and block indices. The variables are replaced with their serialized counterparts. For example, \textit{threadIdx.x} is replaced with \textit{loopIdx} in a single loop, or $\textit{outerIdx} \times \textit{warpSize} + \textit{innerIdx}$ in nested loops. As shown in \Cref{fig:sw:loop}, thread-local variables are converted to arrays, and references to the thread index are replaced with the loop index.

\subsection{Transformation Rules for Warp-level Features}
\label{sec:sw:rule}

\begin{table}[t]
    \centering
    \begin{tabular}{|l|l|}
        \hline
        \textbf{Warp-level Features} & \textbf{Transformation Rule} \\
        \hline
        vote\_any & r = r $\|$ value[tid] \\
        \hline
        vote\_all & r = r \&\& value[tid] \\
        \hline
        vote\_ballot & r = r $|$ ((value[tid] != 0) $\ll$ tid) \\
        \hline
        shuffle & r = value[srcLane] \\
        \hline
        shuffle\_up & r[tid] = value[tid - delta] \\
        \hline
        shuffle\_down & r[tid] = value[tid + delta] \\
        \hline
        shuffle\_xor & r[tid] = value[tid $\oplus$ delta] \\
        \hline  
        \hline
        thread\_group::num\_threads() & group\_size \\
        \hline
        thread\_group::thread\_rank() & tid \% group\_size \\
        \hline
        thread\_group::meta\_group\_rank() & tid / group\_size \\
        \hline
    \end{tabular}
    \caption{PR Transformation rules}
    \vspace{-5mm}
    \label{tab:rules}
\end{table}

\Cref{tab:rules} shows the PR transformation rules for warp-level features during the parallel region transformation.



\textbf{Warp-Level Functions}: To support warp-level functions, a temporary array as large as the warp is constructed to store the function return value of each thread. The return value can be obtained by accessing the array element with the emulated thread index. If a function produces identical results across the warp, the array can be omitted, and a single variable can store the result. \Cref{tab:rules} shows the transformation rules for warp-level functions. This implementation strategy extends to tiles, where the warp size is replaced by the tile size.

\textbf{Cooperative Groups}: While the loop serialization transformation ensures the correct execution order of thread groups, accessor methods, such as the thread rank of the group, can be directly calculated using the tile size and the thread block thread index. \Cref{tab:rules} shows the transformation rules for cooperative group methods.

\section{Evaluation}
\label{sec:eval}



We evaluate our software and hardware solutions by extending the Vortex RISC-V GPU~\cite{vortex, vortex_first_2020}. For performance evaluation, we use the SimX cycle-level simulator of the Vortex GPU~\cite{vortex_first_2020} with six different benchmarks.
The benchmarks consist of two computational kernels (\textit{mse\_forward}~\cite{unet_cuda_repo}, \textit{matmul}), two functionality tests(\textit{shuffle}, and \textit{vote}~\cite{a2023_cuda}) and two reduction kernels (\textit{reduce}, \textit{reduce\_tile}~\cite{a2023_cuda}). For area cost estimation, we synthesize the configuration using Vivado Design Suite 2023.1 for the Xilinx U50 FPGA (xcu50-fsvh2104-2-e).
For both evaluations, the Vortex GPU is configured with eight threads per warp and four warps per thread block for one core.

\subsection{Performance comparisons}
\label{sec:eval:perf}

\begin{figure}[t]
  \centering
  \includegraphics[width=0.95\linewidth]{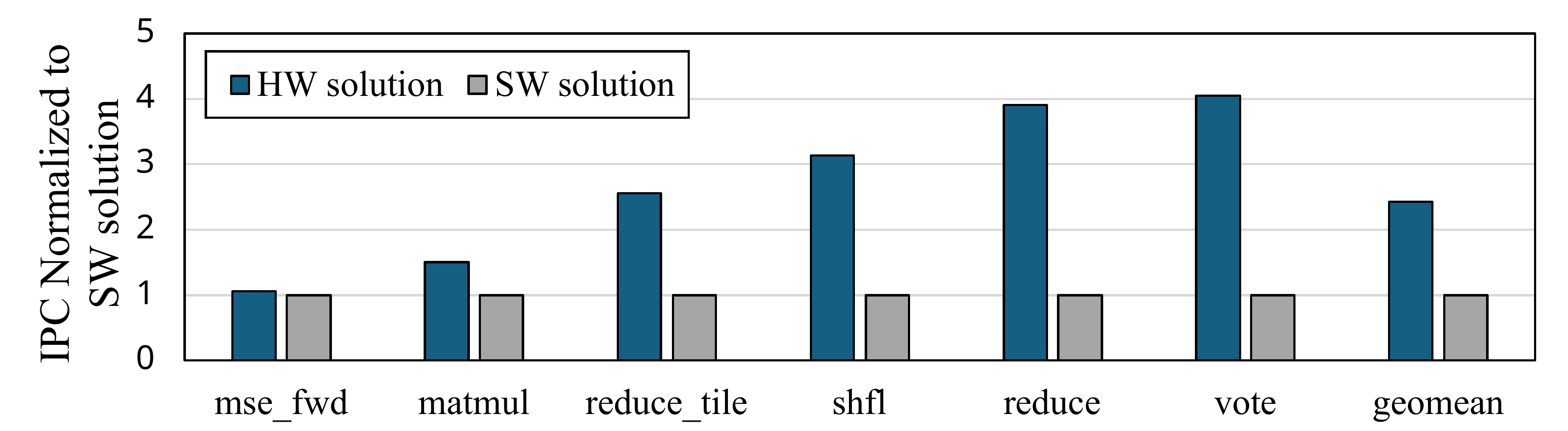}
  \caption{Instruction per Cycle(IPC) of 4 different benchmarks}
\label{fig:eval:perf}
\end{figure}




\Cref{fig:eval:perf} shows the IPC of the HW and SW solutions, with the HW solution achieving a 2.42× geomean speedup compared to the SW solution.
\textit{mse\_forward\_kernel} use shuffle\_down functions, so this kernel with the SW solution benefits from loop serialization by reducing memory access compared to the HW solution.
$\textit{matmul}$ does not involve warp-level collectives, but this kernel exhibits loop serialization overhead in the SW solution, resulting in a 30\% performance loss compared to the HW solution.
The $\textit{vote}$, $\textit{shfl}$ (testing shfl\_sync), \textit{reduce}, and \textit{reduce\_tile} kernels achieve almost 4× speedups with the hardware solution, as the instructions directly access registers instead of using memory.
Even though the HW solution generally shows good performance, the results of \textit{mse\_forward} and \textit{matmul} reveal that software-based solutions can be a viable alternative for some kernels from a performance point of view.






\subsection{Area cost estimation from synthesizations}
\label{sec:eval:area}


\begin{table}[]
    \centering
    \label{tab:Synthesis}
    \begin{tabular}{|c|c|c|} \hline
        {\bf Site Type} & {\bf SLR 0} & {\bf SLR 1}  \\ \hline
        Control Logic Blocks (CLB) & 1.08\% & 0.43\% \\ \hline
        CLB Look-Up Tables (LUTs) &  -0.03\% & 0.00\% \\ \hline
        CLB Registers & 0.25\% & 0.01\% \\ \hline
        Others & -0.26 \% & 0.04\% \\ \hline
        Total Resource Utilization Overhead & 1.04\% & 0.48\%  \\ \hline
    \end{tabular}
     \caption{Resource utilization overhead in Super Logic Regions (SLR) when comparing the change in the Vortex extension with our HW solution against the original Vortex}
     \vspace{-3mm}
\end{table}

\begin{figure}[t]
  \centering
  \begin{subfigure}[b]{0.45\linewidth}
    \centering
    \includegraphics[width=\linewidth]{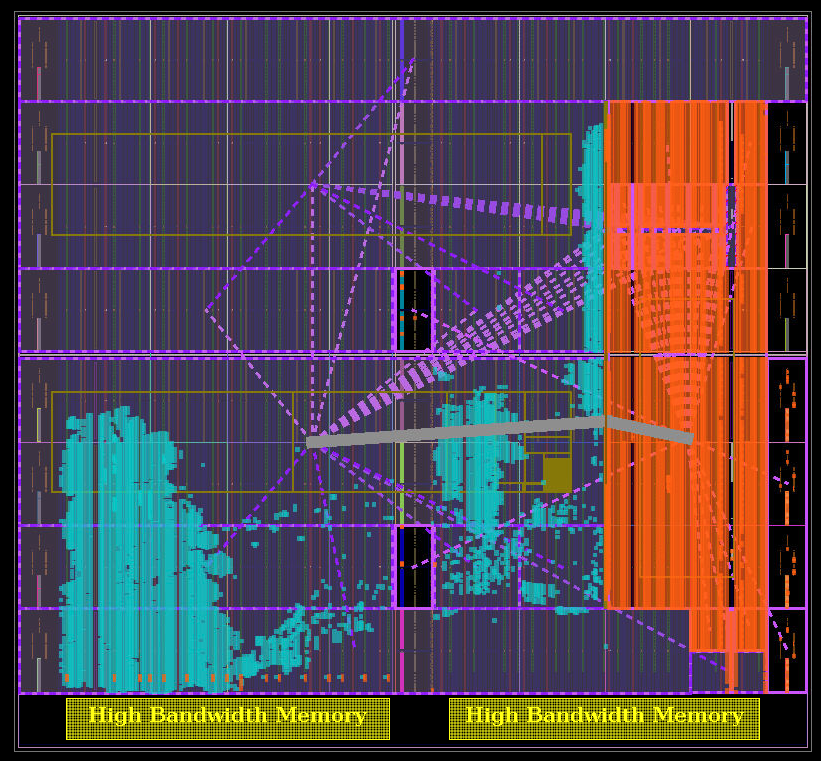}
    \caption{Baseline Design}
    \label{fig:base}
  \end{subfigure}
  \hfill
  \begin{subfigure}[b]{0.45\linewidth}
    \centering
    \includegraphics[width=\linewidth]{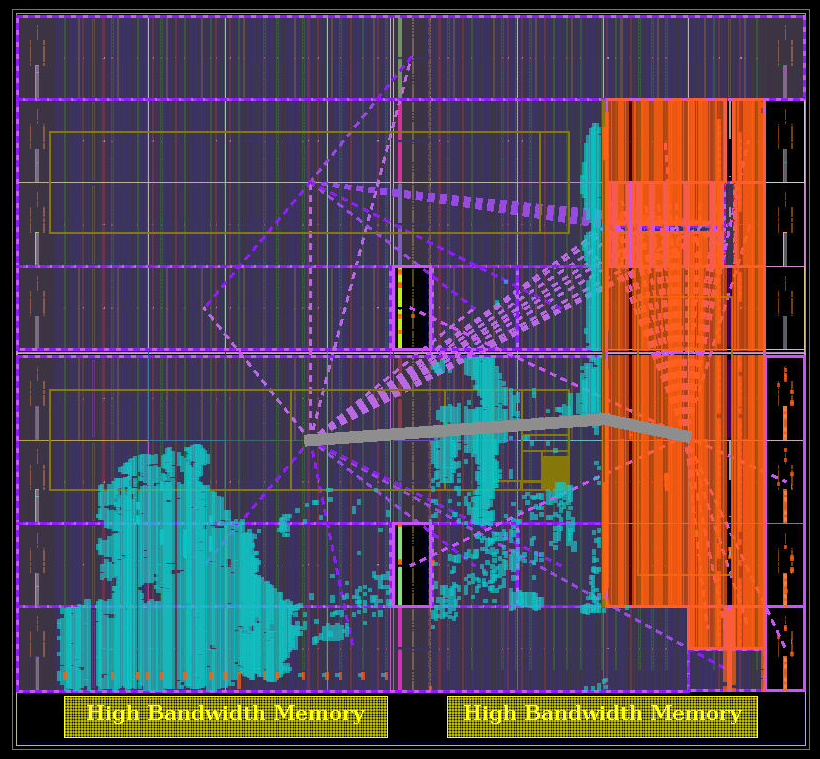}
    \caption{Implemented Design}
    \label{fig:final}
  \end{subfigure}
  \caption{Comparison of Synthesized Layout Figures}
  \label{fig:comparison}
  \vspace{-3mm}
\end{figure}



    Our HW solution synthesized on a Xilinx U50 incurs a total area overhead of approximately 2\% per core. \Cref{fig:comparison} show the synthesized layouts. CLB utilization exhibits the most significant impact by adding warp-level features, increasing by 1.08\% in SLR0 and 0.43\% in SLR1. 
The impact on other resources is negligible due to synthesis optimization variations between designs. In addition, LUT registers show a small increase in both SLR0 and SLR1.

\section{Conclusion}
\label{sec:conclusion}

This paper explores hardware and software approaches to supporting warp-level features and evaluates their trade-offs. The hardware implementation extends the Vortex ISA with minimal overhead, utilizing approximately 2\% of CLB resources per core leading to performance gains for some kernels. While hardware support provides superior performance benefits, software-based solutions are viable when area cost is critical. Therefore, users can select between hardware and software implementations based on application requirements and area constraints. Future work may explore hardware acceleration for complex operations, such as reduction and support for mapping variable CUDA threads to Vortex threads while leveraging hardware features.


{
  \fontsize{7}{8.5}\selectfont
  \setlength{\bibsep}{0pt plus 0.2pt}
  \bibliographystyle{IEEEtran}
  \bibliography{paper}
}

\end{document}